\documentstyle[preprint,aps,graphicx]{revtex}
\tightenlines
\begin{document}
\newcommand{\bref}[1]{eq.~(\ref{#1})}
\newcommand{\be}{\begin{equation}}
\newcommand{\en}{\end{equation}}
\newcommand{\bs}{$\backslash$}
\newcommand{\us}{$\_$}

\title{Faraday rotation and sensitivity of (100) bismuth-substituted 
ferrite garnet films}
\author{L.E. Helseth, A.G. Solovyev and R.W. Hansen}
\address{Department of Physics, University of Oslo, P.O Box 1048 Blindern, N-0316 Oslo, Norway}

\maketitle
\begin{abstract}
We have investigated the Faraday rotation of in-plane magnetized 
bismuth - substituted ferrite garnet films grown by liquid phase epitaxy 
on (100) oriented gadolinium gallium garnet substrates. The Faraday spectra were measured
for photon energies between 1.7 - 2.6 eV. To interprete the spectra, we use
a model based on two electric dipole transitions, one tetrahedral and
one octahedral. Furthermore, the Faraday rotation sensitivity was measured at
2.3 eV, and found to be in good agreement with the theoretical predicitions.
In particular, we find that the sensitivity increases linearly with the
bismuth content and nonlinearly with the gallium content.
 
\end{abstract}

\narrowtext

\newpage

\section{Introduction}
It is well known that bismuth-substituted ferrite garnets (Bi:FGs) have a 
giant magnetooptical response\cite{Wittekoek1,Wittekoek2,Takeuchi,Krinchik,Hansen,Hansen1,Hansen2,Hansen3,Simsa}. 
For this reason, they have found widespread use as optical switchers, optical 
isolators and magnetic field sensors. Some years ago it was realized 
that Bi:FG films with in-plane magnetization allow effective visualization 
and detection of magnetic fields\cite{Belyaeva,Dorosinskii}. This 
discovery triggered a large number of quantitative studies of magnetic fields 
from superconductors, domain formation in magnetic materials, currents in microelectronic circuits 
and recorded patterns in magnetic storage
media\cite{Koblischka,Vlasko-Vlasov,Hubert,Egorov,Kotov,Shamonin}. 
However, no systematic studies have been carried out to characterise the
magnetooptic properties of these films. Of particular interest
here is the Faraday rotation, since this parameter determines the 
usefulness of the indicator.

Bi:FG films grown on (100) oriented substrates have been shown to have a 
number of unique properties which make them excellent candidates for 
magnetooptic imaging. First, these films exihibit very little domain activity, 
and respond to an increasing external field by a continous rotation of the 
magnetization vector. Second, the sensitivity of the films is easily tuned by 
altering the chemical composition. The sensitivity of the Faraday rotation to an external field 
is of major importance in magnetooptic imaging and detection, in particular 
when the external field is weak. To date several studies have been done to 
determine the sensitivity of bulk-samples and (111) oriented Bi:FG
films\cite{Deeter1,Deeter2,Deeter3,Syvorotka,Wallenhorst}. However, no 
studies have been carried out to characterise and understand the sensitivity 
of (100) orientated films.

In a previous paper we presented an experimental and theoretical study of the 
Faraday rotation at saturation\cite{Helseth}. In that study a very simple 
model based on two electric dipole transitions, one tetrahedral and one 
octahedral, was used to explain the experimental data. Here we extend the 
work in that paper, and also introduce several new features. We have grown 
a series of gallium-substituted Bi:FG films using the liquid phase epitaxy 
(LPE) technique, and characterised their chemical composition. The Faraday 
rotation spectra have been measured for photon energies between $1.7$ and 
$2.6$ eV, corresponding to wavelengths between $730$ and $480$ nm. It is 
shown that the Faraday rotation changes significantly with 
the amount of substituted gallium and bismuth. Furthermore, the comparison 
of experimental and theoretical data confirms that the magnetooptic response 
increases linearly with the bismuth substitution and decreases almost 
linearly with the gallium substitution. We also report experimental data for 
the sensitivity of these materials, and find that the sensitivity depends nonlinearly on the gallium
substitution, in good agreement with the theoretical model.

\section{Sample preparation}
Single crystal films of Bi:FG were grown by isothermal LPE from
$Bi_{2}O_{3}/PbO/B_{2}O_{3}$ flux onto (100) oriented gadolinium gallium
garnet (GGG) substrates. The growth takes place while the  
substrate is dipped into the melt contained in a Pt crucible. During the 
growth, the parameters could be controlled to create low magnetic coercivity 
and in-plane magnetization in the garnet films. The thickness 
of the films was measured using a Scanning Electron Microscope (SEM) and
confirmed with optical techniques, while their composition were determined 
with an Electron MicroProbe (EMP). Thicknesses and compositions of the 
selected samples are listed in Table 1. 

From Table I one sees that the films can be represented by the following 
general formula: $\{Re_{3-x}Bi_{x}\}[Fe_{2-z_{a}}Ga_{z_{a}}](Fe_{3-z_{d}}Ga_{z_{d}})O_{12}$,
where $\{ \}$ indicates the dodecahedral site, [ ] the octahedral site, and () the
tetrahedral site. Note that only the total gallium content $z=z_{a}+z_{d}$ can be
extracted from the EMP, and to determine $z_{a}$ and $z_{d}$ separately other 
techniques such as neutron spectroscopy must be applied. The distribution of
gallium on tetrahedral and octahedral sites have been examined in a number of
studies on garnets with and without bismuth\cite{Czerlinsky,Scott,Fratello}. 
These results indicate that around $90\%$ of the gallium occupies the
tetrahedral site. This fact will be used later in this paper.  

The films also contain small amounts of $Pb$, typically of the order 
of $0.05$, but this will neglected here, since it does not influence the
Faraday rotation significantly. 

\section{Faraday rotation spectra}
We have measured the Faraday spectra (at room temperature) of the films 
presented in Table I. Shown in Figs. \ref{f1} and \ref{f2} is the observed Faraday rotation as a function 
of wavelength for samples 7 and 8, respectively. Note that both films
have maximum rotation near 2.45 eV, a feature characteristic of all the films
considered in this study. In fact, all the films exhibit the same spectral
shape, with only minor deviations from that seen in Figs. \ref{f1} and
\ref{f2}. 

In order to understand the behaviour of the spectra, we have adopted the 
theory developed in Refs.\cite{Helseth,Dionne,Dionne1}. Here the expression for 
the Faraday rotation is given by
\begin{eqnarray} 
\Theta_{F} ^{sat} &=& \frac{\pi e^{2} \omega ^{2}}{nmc} \sum_{i=a,d}\frac{Nf_{i}}{\omega_{i}} \nonumber \\
& & \times\left\{ 
\frac{(\omega_{i}+\Delta_{i})^{2}-\omega^{2}-\Gamma_{i}^{2}}{\left[(\omega_{i}+\Delta_{i})^{2}-\omega^{2}+\Gamma_{i}^{2}
\right]^{2}
+ 4\omega^{2}\Gamma_{i}^{2}} -
\frac{(\omega_{i}-\Delta_{i})^{2}-\omega^{2}-\Gamma_{i}^{2}}{\left[(\omega_{i}-\Delta_{i})^{2}-\omega^{2}+\Gamma_{i}^{2}\right]^{2}
+ 4\omega^{2}\Gamma_{i}^{2}} \right\} \,\,\,  ,
\label{Far} 
\end{eqnarray}
where $\omega_{i}$ represent the resonance energy, $f_{i}$ the oscillator 
strength, while $\Gamma_{i}$ is the half-linewidth of the transition. 
Furthermore, $e$ and $m$ are the electron charge and mass, respectively, 
whereas $N$ is the active ion density.

For $x<2$ it is reasonable to assume that $N$ is directly
proportional to the bismuth content $x$\cite{Helseth}. Furthermore, it is 
known that the strong enhancement of Faraday rotation is caused by iron-pair 
transitions, involving both octahedral and tetrahedral transitions 
simultaneously\cite{Scott}. Therefore, iron dilution of either sublattice 
results in a reduction of the active ion density. For these reasons we 
assume that the active ion density can be written as\cite{Helseth}
\begin{equation}
N=N_{0}(1-z_{d}/3)(1-z_{a}/2)x  \,\,\, .
\label{ion}
\end{equation}
$N_{0}$ is a constant, and may expected to be 1/3 of the density of rare-earth ions on the 
dodecahedral site, i.e. $1.3\times10^{22}$ $cm^{-3}/3$. When $x=3$, this
interpretation implies that the dodecahedral site is fully occupied by bismuth.

To fit theoretical curves to the experimental data, the product $N_{0}f_{i}$
was chosen as free parameter. The parameters $\Delta _{i}$, $\omega _{i}$ and
$\Gamma _{i}$ were chosen as sample independent, and the values suggested in
Ref. \cite{Helseth} were used as a starting point in the fitting. 
Table II presents the paremeters found to give the best fit between the
theoretical curves and experimental data. Note that our values for 
$\Delta _{i}$, $\omega _{i}$ and $\Gamma _{i}$ differs slightly from those
used in Ref. \cite{Helseth}. This is due to the fact that in that paper we
focused on obtaining a very good fit for energies less than 2.3 eV. However,
this resulted in some deviations between experiment and theory near the 
maximum Faraday rotation. Here we have chosen parameters which give 
the best agreement within the whole range of experimental data. 

The solid line shown in Fig. \ref{f1} is the Faraday rotation calculated from 
Eq. (\ref{Far}) using the values in Table II. The dashed and dash-dotted lines
show the contribution to the total Faraday rotation from the octahedral and
tetrahedral sites, respectively. Note that below 2.2 eV, the main contribution
to $\Theta _{F}$ comes from the octahedral site.

The solid line in Fig. \ref{f2} shows the calculated Faraday rotation
of sample 8. Again, we note that the theoretical curve is in good agreement 
with the experimental data. Also shown is the theoretical predictions when 
the bismuth content is x=1.5 (dashed line). In this case one may expect 
almost 7 $deg/\mu m$ at 2.3 eV.

It is useful to find out how bismuth and gallium influences the Faraday
rotation. To this end, we define the following parameters
\begin{equation}
\Theta _{F} ^{Bi} =\frac{\Theta_{F} ^{sat}}{(1-z_{d}/3)(1-z_{a}/2)} \,\,\,  ,
\end{equation} 
and 
\begin{equation}
\Theta _{F} ^{Ga} =\frac{\Theta_{F} ^{sat}}{x} \,\,\,  .
\end{equation} 
Here $\Theta _{F} ^{Bi}$ and $\Theta _{F} ^{Ga}$ are the Faraday rotations 
associated with the bismuth and gallium content, respectively. The
experimental data and theoretical curves for these two parameters are plotted
in Figs. \ref{f3} and \ref{f4}, respectively. The good agreement between
experimental data and theoretical predictions confirm the validity of the
model used here.

\section{Sensitivity}
If a light beam propagates along the z-axis through the magnetic film, then 
the polar Faraday rotation of the film is given by (neglecting the Voigt-effect and 
multiple reflections)\cite{Shamonin} 
\begin{equation}
\Theta _{F} =\Theta _{F} ^{sat} \frac{H_{z}}{H_{a}}, \,\,\,\, H_{z} \leq H_{a}\,\,\, , 
\label{F}
\end{equation}
where the anisotropy field is defined by
\begin{equation}
H_{a}=M_{s} - \frac{2K_{u}^{tot}}{\mu _{0}M_{s}}\,\,\, .
\end{equation}
Eq. (\ref{F}) is only valid as long as the cubic anisotropy can be neglected.
When $H_{z} \geq H_{a}$, the Faraday rotation is at its maximum value. As an 
example, the Faraday rotation as a function of $H_{z}$ for sample 4 is shown 
in Fig. \ref{f5}. Note that the Faraday rotation is accurately described by
Eq. (\ref{F}) as long as $H_{z} < H_{a}$. In Fig. \ref{f6} the experimental 
values for $H_{a}$ is displayed as a function of gallium substitition. The dashed line
shows the best fit to $H_{a}$ for films with composition 
$Lu_{3-x}Bi_{x}Fe_{5-z}Ga_{z}O_{12}$
\begin{equation}
H_{a} \approx 210000(1-0.7z)  \,\,\, .
\label{Ha}
\end{equation}
Also shown is the value for $M_{s}$ 
found in the litterature\cite{Hansen,Hansen1,Fratello}
\begin{equation}
M_{s} \approx 160000(1-0.75z)  \,\,\, .
\label{Ms}
\end{equation}  
The experimental values for $H_{a}$ are quite close to Eq. (\ref{Ms}), which 
indicates that the uniaxial anisotropy plays a minor role here. Since the films of composition $Lu_{3-x}Bi_{x}Fe_{5-z}Ga_{z}O_{12}$ are well
described by Eq. (\ref{Ha}), we will use this in the further modelling. 
Film 7, 8 and 12 are not well described by Eq. (\ref{Ha}), which is most 
probably due to their chemical composition.  

The Faraday rotation sensitivity is given by 
\begin{equation}
S = \frac{d\Theta _{F}}{dH_{z}} =\frac{\Theta _{F} ^{sat} }{H_a}  \,\,\, .
\label{sens}
\end{equation} 
It is useful to separate the contributions from bismuth and gallium.
The contribution from bismuth can be written as
\begin{equation}
S^{Bi} =\frac{S(1-0.7z)}{(1-z_{d}/3)(1-z_{a}/2)}  \,\,\, .
\label{BI}
\end{equation}
The experimental data at 2.3 eV (wavelength 540 nm) are shown in 
Fig. \ref{f7} together with the theoretical prediction based on 
Eq. (\ref{BI}). A reasonably good agreement is 
obtained for all samples of composition $Lu_{3-x}Bi_{x}Fe_{5-z}Ga_{z}O_{12}$, 
except sample 11, which shows a minor deviation from the theoretical curve. We
do not know the reason for this deviation. 
It is interesting to observe that the sensitivity depends linearly on 
the bismuth content.
The experimental data for samples 7, 8 and 12 (diamond, circle and cross) are 
located far from the straight line, since they do not follow Eq. (\ref{Ha}).

It is also of interest to see how gallium influences the sensitivity of the
material. To that end, we define
\begin{equation}
S^{Ga} =\frac{S}{x}  \,\,\, .
\label{GA}
\end{equation}
In Fig. \ref{f8} the experimental data are shown together with the
theoretical prediction based on Eq. (\ref{GA}). Again
we note that there is good agreement between the experimental data and the 
theoretical curve for samples of composition 
$Lu_{3-x}Bi_{x}Fe_{5-z}Ga_{z}O_{12}$ (except for sample 11, which shows a
minor deviation from the theoretical curve). Now the sensitivity
has a strong nonlinear dependence on the gallium content. Thus, one may think that
it should be possible to increase the sensitivity even further by 
adding more gallium. However, this is not the case. If we add more gallium 
(above z=1.2), the film approaches its compensation point ($M_{s}=0$) where 
the coercivity of the material is rather high. On the other hand, it would 
be of interest to investigate the sensitivity above the compensation point,
which is outside the scope of this paper.    

\section{Conclusion}
We have investigated the Faraday rotation of in-plane magnetized 
bismuth - substituted ferrite garnet films grown by liquid phase epitaxy 
on (100) oriented gadolinium gallium garnet substrates. 
The Faraday spectra were measured for photon energies between 1.7 - 2.6 eV. 
To interprete the spectra, we use a simple model based on two electric dipole 
transitions, and find excellent agreement with the experimental data. 
Furthermore, the Faraday rotation sensitivity was measured at 2.3 eV, and 
found to be in good agreement with the theoretical predicitions.
In particular, we find that the sensitivity increases linearly with the
bismuth content and nonlinearly with the gallium content.

\acknowledgements
The authors are grateful to E.I. Il'yashenko for supplying the 
GGG substrates and for helpful discussions. This work was financially 
supported by The Norwegian Research Council and Tandberg Data ASA.

\newpage

$\bf{Table}$ $\bf{I}$: The thickness and chemical composition of the  
samples.
\\
\\
\\
\begin{tabular}{c c c c c c c c}
\hline \hline
Sample nr.\,\,\,\,\,\,\,\,\, 
& Lu\,\,\,\,\,\,\,\,\,
& Y\,\,\,\,\,\,\,\,\, 
& Tm\,\,\,\,\,\,\,\,\, 
& Bi\,\,\,\,\,\,\,\,\,   
& Fe\,\,\,\,\,\,\,\,\,
& Ga\,\,\,\,\,\,\,\,\, 
& t ($\mu m$) \\
\hline
1\,\,\,\,\,\,\,\,\,
& 2.5\,\,\,\,\,\,\,\,\, 
& 0\,\,\,\,\,\,\,\,\, 
& 0\,\,\,\,\,\,\,\,\,  
& 0.5\,\,\,\,\,\,\,\,\, 
& 4.9\,\,\,\,\,\,\,\,\,  
& 0.1\,\,\,\,\,\,\,\,\,
& 4.0   
\\

2\,\,\,\,\,\,\,\,\,
& 2.4\,\,\,\,\,\,\,\,\,  
& 0\,\,\,\,\,\,\,\,\,  
& 0\,\,\,\,\,\,\,\,\, 
& 0.6\,\,\,\,\,\,\,\,\, 
& 4.8\,\,\,\,\,\,\,\,\,  
& 0.2\,\,\,\,\,\,\,\,\,   
& 3.5
\\

3\,\,\,\,\,\,\,\,\, 
& 2.3\,\,\,\,\,\,\,\,\,  
& 0\,\,\,\,\,\,\,\,\,  
& 0\,\,\,\,\,\,\,\,\,  
& 0.7\,\,\,\,\,\,\,\,\, 
& 4.7\,\,\,\,\,\,\,\,\,  
& 0.3\,\,\,\,\,\,\,\,\,  
& 3.5

\\
4\,\,\,\,\,\,\,\,\, 
& 2.3\,\,\,\,\,\,\,\,\,    
& 0\,\,\,\,\,\,\,\,\,  
& 0\,\,\,\,\,\,\,\,\, 
& 0.7\,\,\,\,\,\,\,\,\,  
& 4.4\,\,\,\,\,\,\,\,\, 
& 0.6\,\,\,\,\,\,\,\,\, 
& 4.0 \\

5\,\,\,\,\,\,\,\,\, 
& 2.3\,\,\,\,\,\,\,\,\,    
& 0\,\,\,\,\,\,\,\,\,  
& 0\,\,\,\,\,\,\,\,\, 
& 0.7\,\,\,\,\,\,\,\,\,  
& 4.2\,\,\,\,\,\,\,\,\, 
& 0.8\,\,\,\,\,\,\,\,\, 
& 4.0 \\

6\,\,\,\,\,\,\,\,\, 
& 2.4\,\,\,\,\,\,\,\,\,    
& 0\,\,\,\,\,\,\,\,\,  
& 0\,\,\,\,\,\,\,\,\, 
& 0.6\,\,\,\,\,\,\,\,\,  
& 4.1\,\,\,\,\,\,\,\,\, 
& 0.9\,\,\,\,\,\,\,\,\, 
& 3.3 \\

7\,\,\,\,\,\,\,\,\, 
& 0\,\,\,\,\,\,\,\,\,    
& 0\,\,\,\,\,\,\,\,\,  
& 2.3\,\,\,\,\,\,\,\,\, 
& 0.7\,\,\,\,\,\,\,\,\,  
& 4.1\,\,\,\,\,\,\,\,\, 
& 0.9\,\,\,\,\,\,\,\,\, 
& 3.5 \\

8\,\,\,\,\,\,\,\,\, 
& 1.4\,\,\,\,\,\,\,\,\,    
& 1\,\,\,\,\,\,\,\,\,  
& 0\,\,\,\,\,\,\,\,\, 
& 0.6\,\,\,\,\,\,\,\,\,  
& 4.1\,\,\,\,\,\,\,\,\, 
& 0.9\,\,\,\,\,\,\,\,\, 
& 4.0 \\

9\,\,\,\,\,\,\,\,\, 
& 2.2\,\,\,\,\,\,\,\,\,    
& 0\,\,\,\,\,\,\,\,\,  
& 0\,\,\,\,\,\,\,\,\, 
& 0.8\,\,\,\,\,\,\,\,\,  
& 3.8\,\,\,\,\,\,\,\,\, 
& 1.2\,\,\,\,\,\,\,\,\, 
& 2.6 \\

10\,\,\,\,\,\,\,\,\,   
& 2.2\,\,\,\,\,\,\,\,\,  
& 0\,\,\,\,\,\,\,\,\,  
& 0\,\,\,\,\,\,\,\,\, 
& 0.8\,\,\,\,\,\,\,\,\,  
& 3.9\,\,\,\,\,\,\,\,\, 
& 1.1\,\,\,\,\,\,\,\,\, 
& 4.0 \\

11\,\,\,\,\,\,\,\,\, 
& 2.1\,\,\,\,\,\,\,\,\,    
& 0\,\,\,\,\,\,\,\,\,  
& 0\,\,\,\,\,\,\,\,\, 
& 0.9\,\,\,\,\,\,\,\,\,  
& 3\,\,\,\,\,\,\,\,\, 
& 1\,\,\,\,\,\,\,\,\, 
& 4.0 \\

12\,\,\,\,\,\,\,\,\, 
& 1.4\,\,\,\,\,\,\,\,\,    
& 0.7\,\,\,\,\,\,\,\,\,  
& 0\,\,\,\,\,\,\,\,\, 
& 0.7\,\,\,\,\,\,\,\,\,  
& 3.8\,\,\,\,\,\,\,\,\, 
& 1.2\,\,\,\,\,\,\,\,\, 
& 7.5 \\
\hline \hline
\end{tabular} \\

\newpage
$\bf{Table}$ $\bf{II}$: The parameters found to give the best fit between  
Eq. (\ref{Far}) and the experimental data. Note that the tetrahedral and 
octahedral sites are given different signs, since they contribute oppositely 
to the Faraday rotation.
\\
\\
\\

\begin{tabular}{c c c c c}
\hline \hline
site\,\,\,\, 
& $N_{0}f_{i}$ ($cm^{-3}$)    \,\,\,\,
& $\Delta_{i}$ ($eV$)\,\,\,\,
& $\omega_{i}$ ($eV$)\,\,\,\,  
& $\Gamma_{i}$ ($eV$)\\
\hline
a\,\,\,\,
& $2.2*10^{23}$\,\,\,\, 
& $0.4$ \,\,\,\,  
& $3.10$ \,\,\,\,
& $0.5$ \\

d\,\,\,\,
&$-6.4*10^{22}$\,\,\,\,  
&$0.1$ \,\,\,\,
&$2.47$\,\,\,\,
&$0.3$\\
\hline \hline
\end{tabular} \\

\newpage

\begin{figure}
\caption{The Faraday rotation as a function of wavelength for sample 7. The
dashed line is the contribution from the octahedral transition, whereas the
dash-dotted line is due to the tetrahedral contribution.  
\label{f1}}
\end{figure}

\begin{figure}
\caption{The Faraday rotation as a function of wavelength for sample 8. The 
dashed line shows the predicted Faraday rotation for x=1.5 and z=0.9. 
\label{f2}}
\end{figure}

\begin{figure}
\caption{$\Theta_{F}^{Bi}$ ($deg/\mu m$) as a function of bismuth substitution at 2.3 eV.
The solid line is the theoretical curve obtained using Eq. \ref{Far}. 
\label{f3}}
\end{figure}

\begin{figure}
\caption{$\Theta_{F}^{Ga}$ ($deg/\mu m$) as a function of gallium substitution at 2.3 eV.
 The solid line is the theoretical curve obtained using Eq. \ref{Far}.
\label{f4}}
\end{figure}

\begin{figure}
\caption{The Faraday rotation as a function of $H_{z}$ for sample 4 for a 
photon energy of 2.3 eV. 
\label{f5}}
\end{figure}

\begin{figure}
\caption{The saturation field as a function of gallium substitution. 
\label{f6}}
\end{figure}

\begin{figure}
\caption{The sensitivity as a function of bismuth substitution for a photon
energy of 2.3 eV. 
\label{f7}}
\end{figure}

\begin{figure}
\caption{The sensitivity as a function of gallium substitution for a photon
energy of 2.3 eV. 
\label{f8}}
\end{figure}

\newpage
\centerline{\includegraphics[width=14cm]{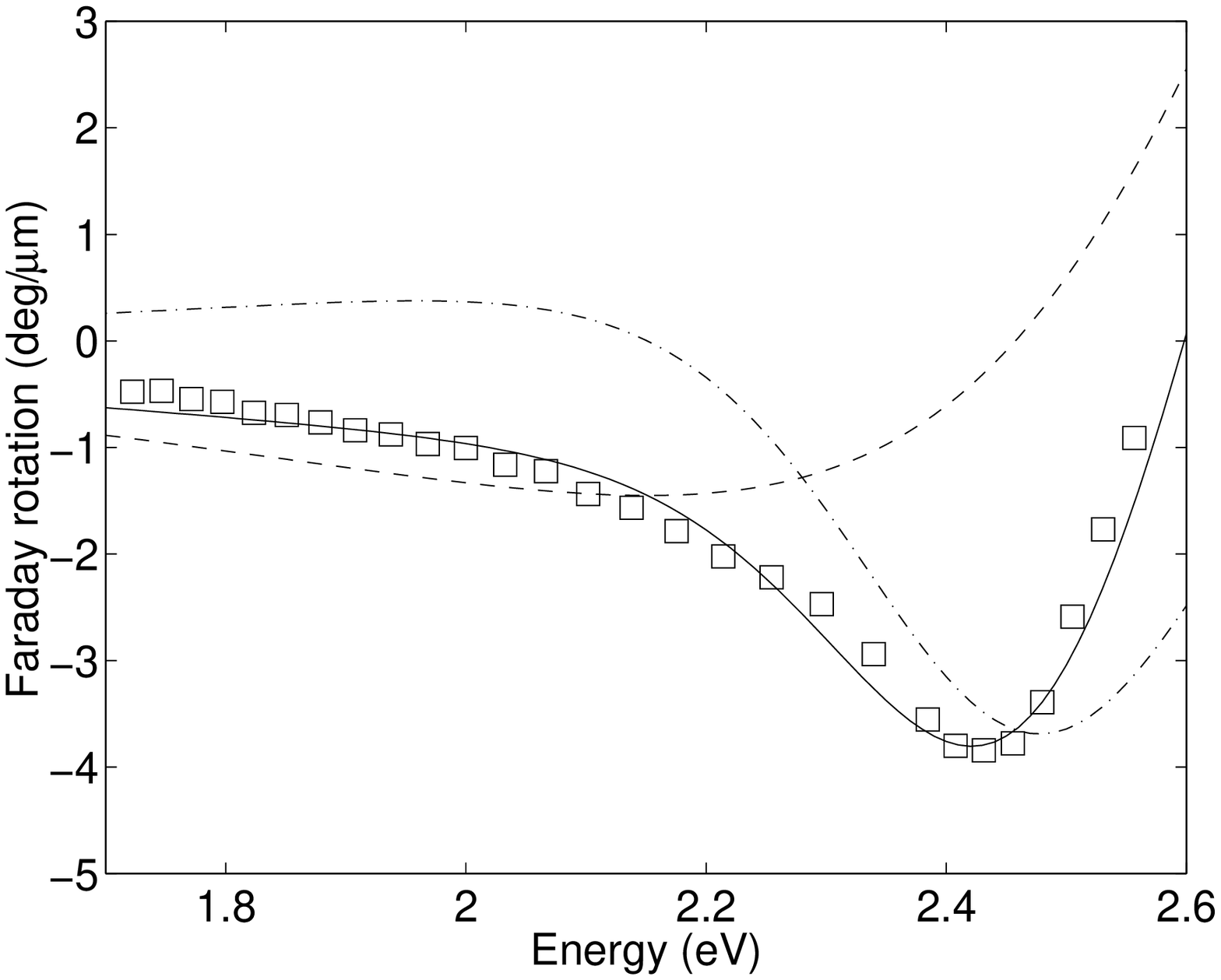}}
\vspace{2cm}
\centerline{Figure~\ref{f1}}

\newpage
\centerline{\includegraphics[width=14cm]{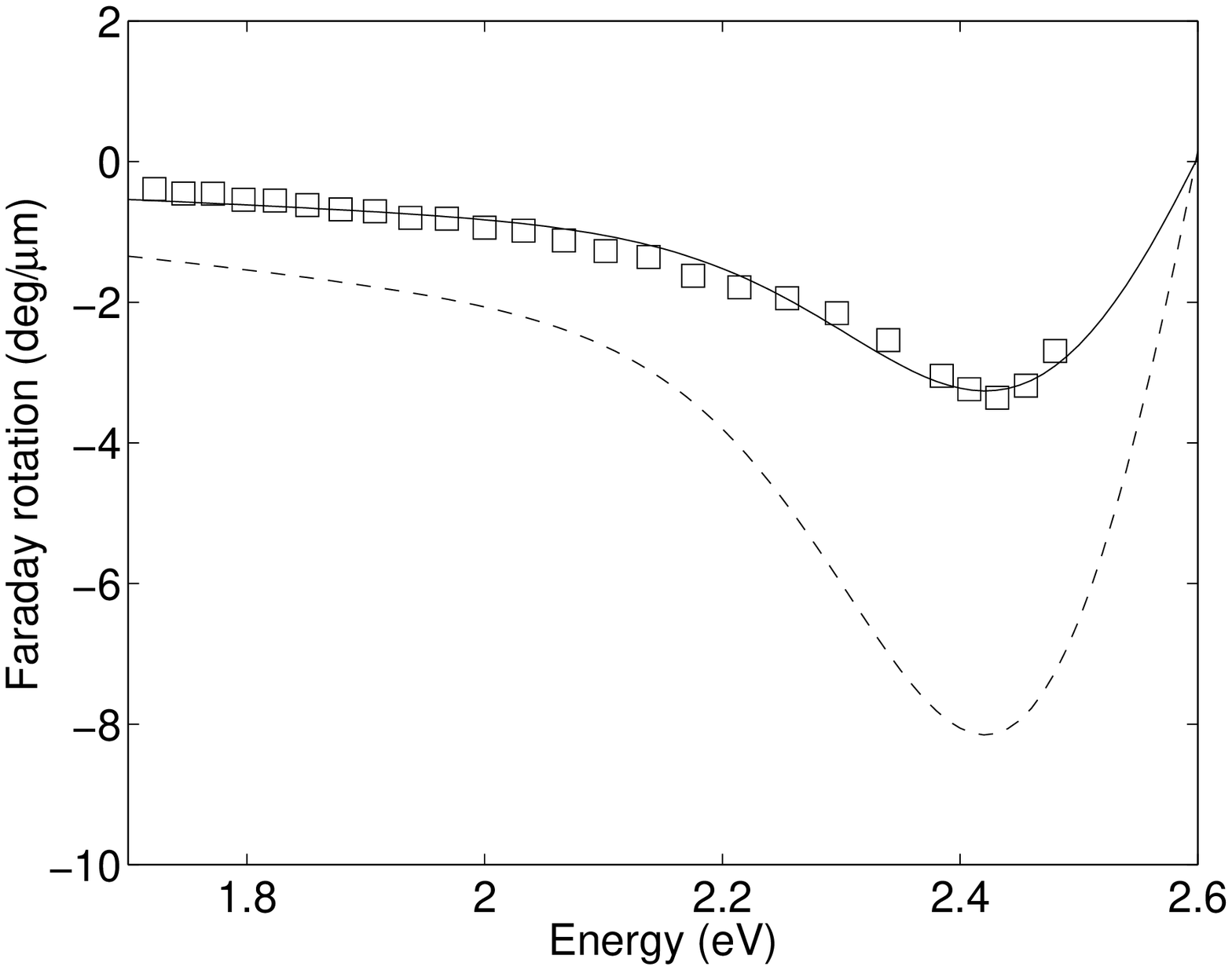}}
\vspace{2cm}
\centerline{Figure~\ref{f2}}

\newpage
\centerline{\includegraphics[width=14cm]{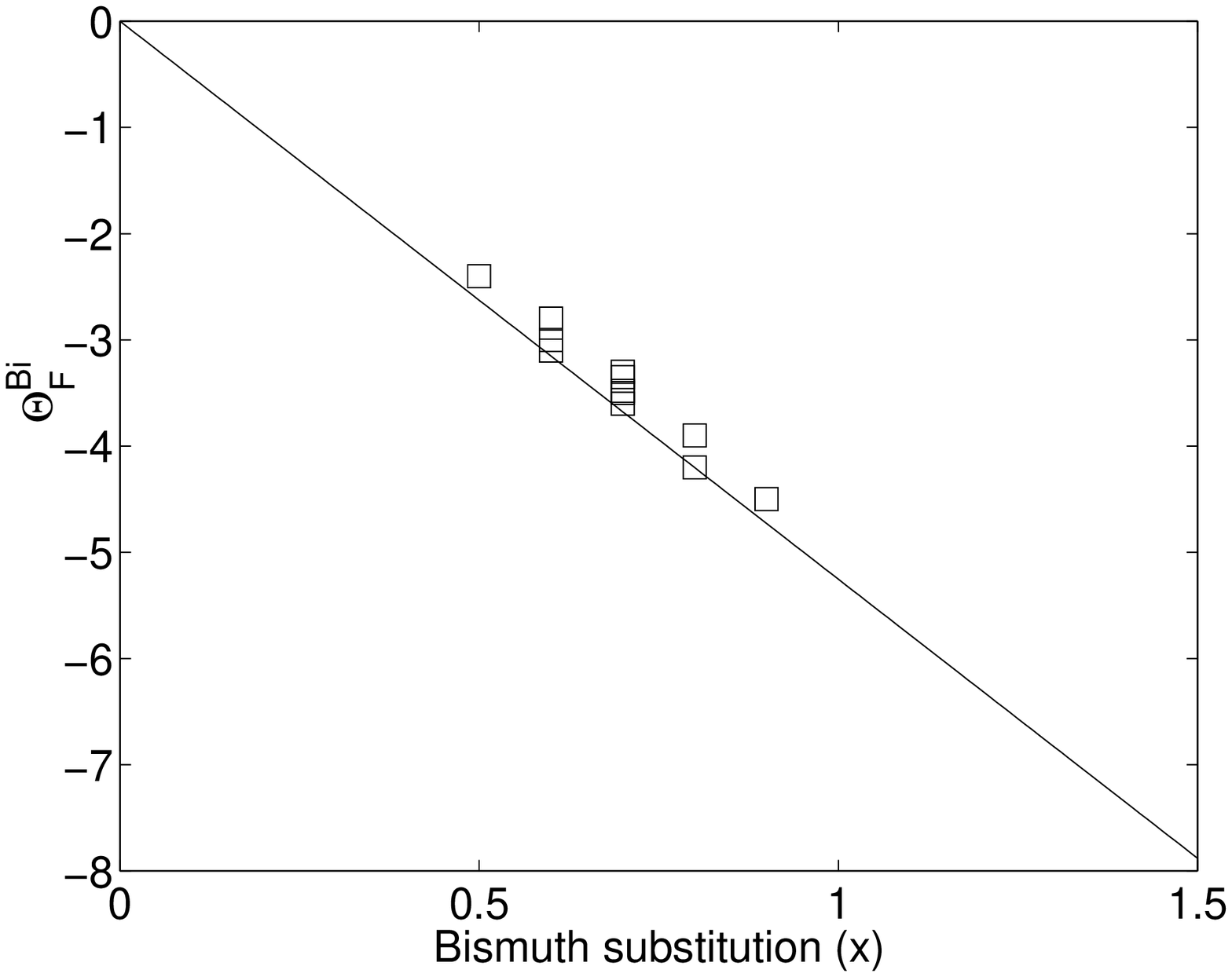}}
\vspace{2cm}
\centerline{Figure~\ref{f3}}

\newpage
\centerline{\includegraphics[width=14cm]{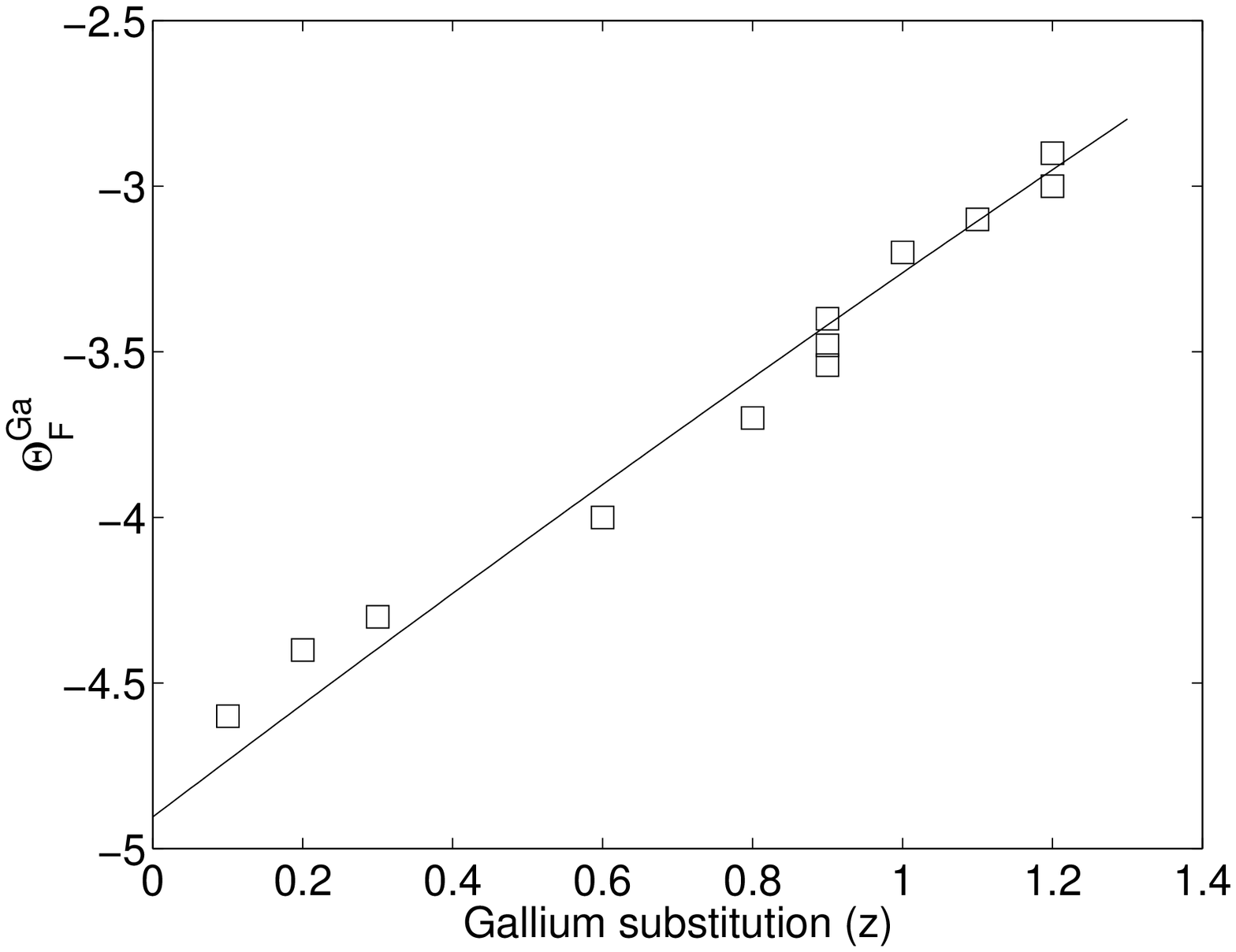}}
\vspace{2cm}
\centerline{Figure~\ref{f4}}

\newpage
\centerline{\includegraphics[width=14cm]{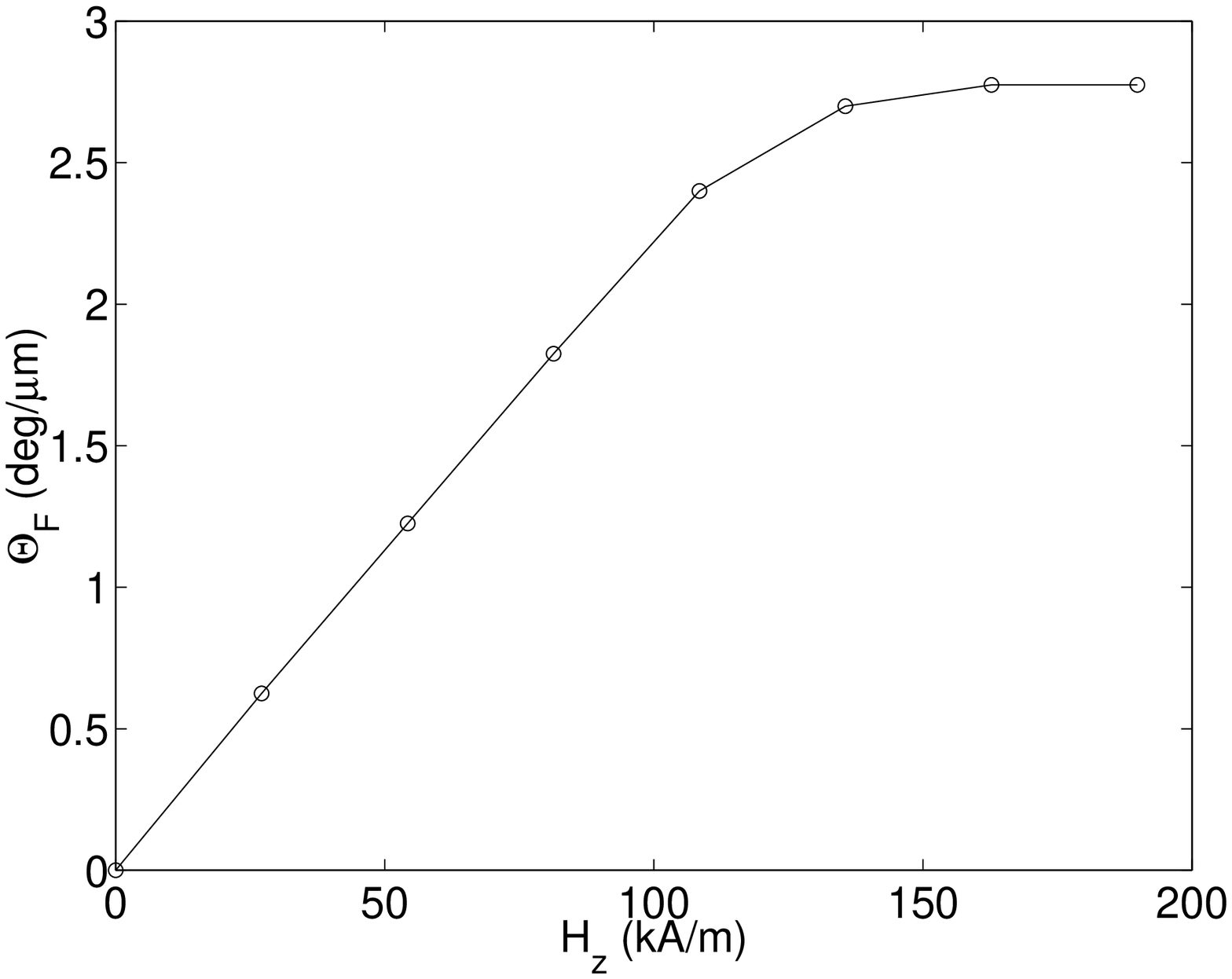}}
\vspace{2cm}
\centerline{Figure~\ref{f5}}

\newpage
\centerline{\includegraphics[width=14cm]{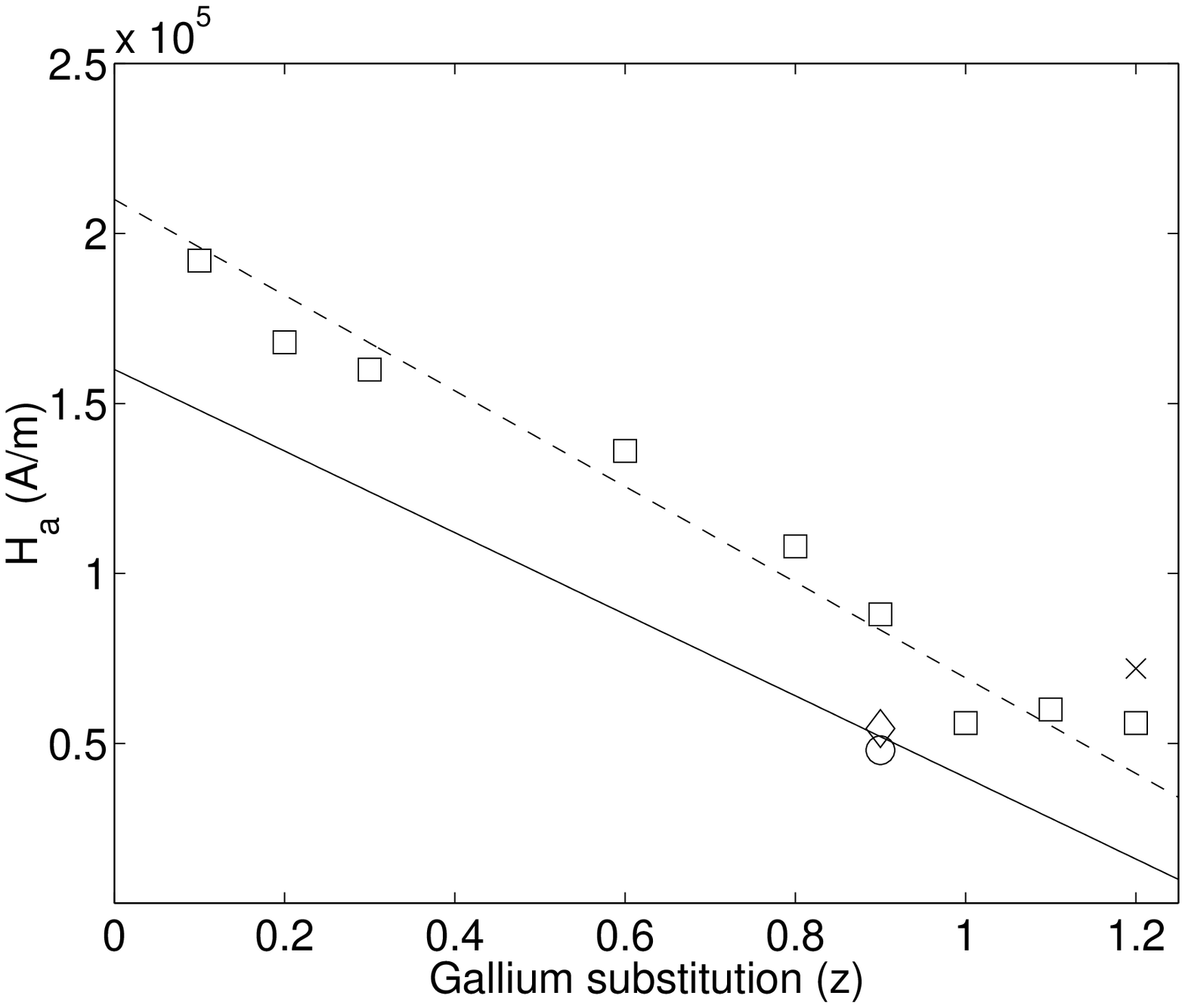}}
\vspace{2cm}
\centerline{Figure~\ref{f6}}

\newpage
\centerline{\includegraphics[width=14cm]{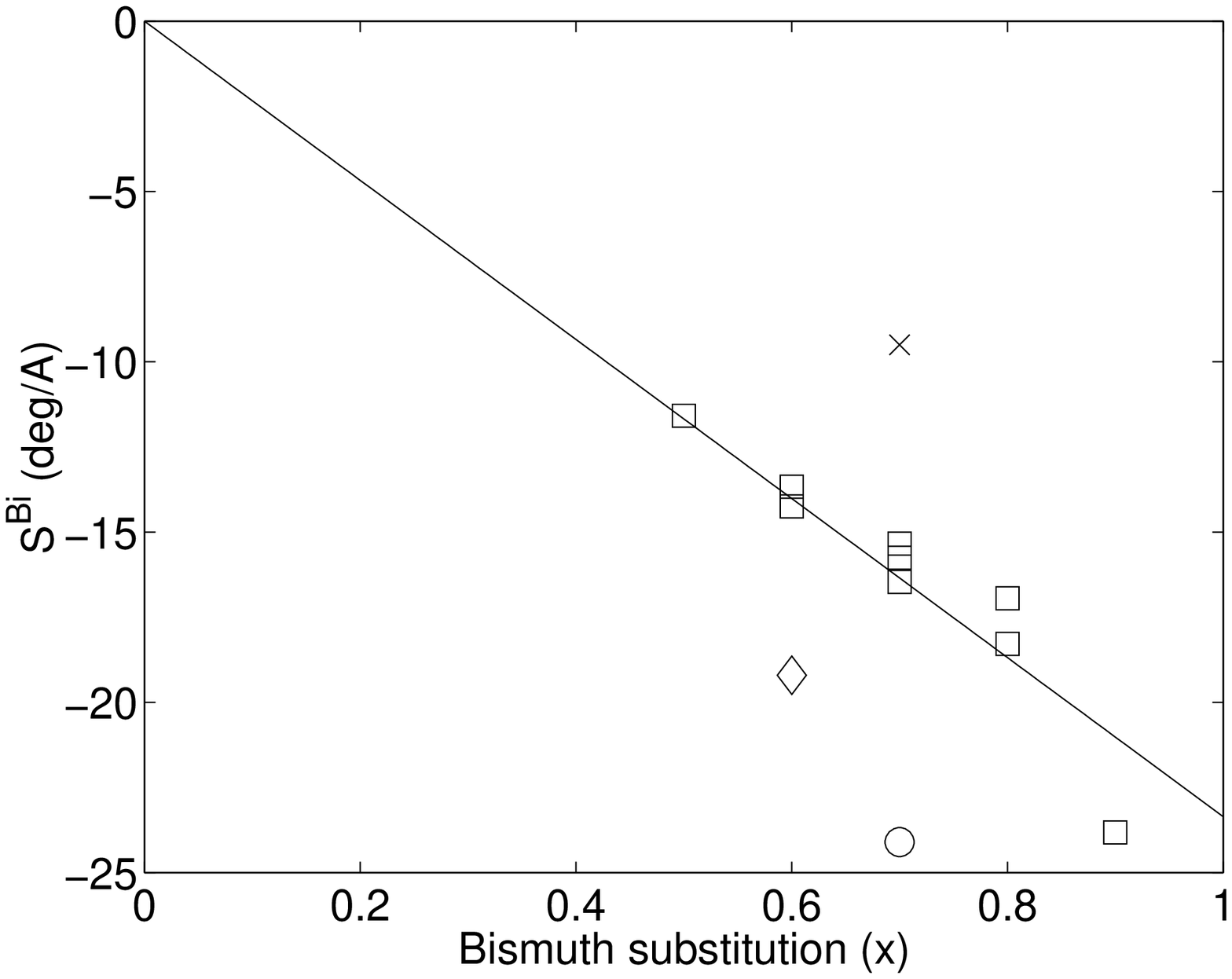}}
\vspace{2cm}
\centerline{Figure~\ref{f7}}

\newpage
\centerline{\includegraphics[width=14cm]{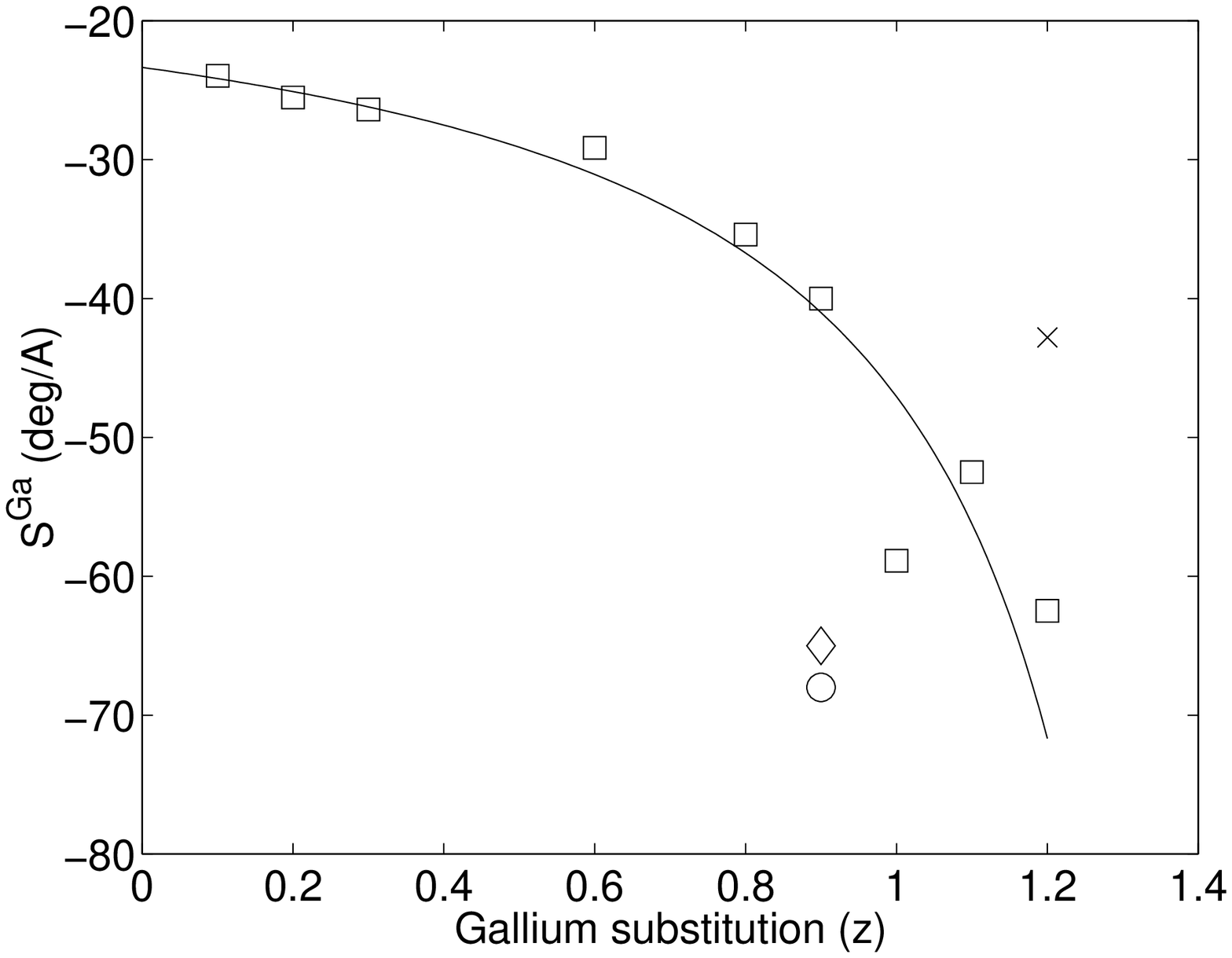}}
\vspace{2cm}
\centerline{Figure~\ref{f8}}

\end{document}